\documentstyle[twoside,fleqn,espcrc2,psfig]{article}

\newcommand{\AmS}{{\protect\the\textfont2
  A\kern-.1667em\lower.5ex\hbox{M}\kern-.125emS}}
\newcommand{\be}{\begin{equation}}
\newcommand{\ee}{\end{equation}}
\def\reff#1{(\ref{#1})}

\newcommand{\1}{1\!\!\!\bot}
\newcommand{\ba}{\begin{eqnarray}}
\newcommand{\ea}{\end{eqnarray}}
\newcommand{\tr}{\mathop{\rm Tr}\nolimits}
\def\spose#1{\hbox to 0pt{#1\hss}}
\def\ltapprox{\mathrel{\spose{\lower 3pt\hbox{$\mathchar"218$}}
 \raise 2.0pt\hbox{$\mathchar"13C$}}}
\def\gtapprox{\mathrel{\spose{\lower 3pt\hbox{$\mathchar"218$}}
 \raise 2.0pt\hbox{$\mathchar"13E$}}}
\title{1) The Influence of Gribov Copies on Gluon and Ghost Propagators
       in Landau Gauge and
       2) A New Implementation
       of the Fourier Acceleration Method}

\author{Attilio Cucchieri 
        and Tereza Mendes\address{Gruppo APE,
        Dipartimento di Fisica,
        Universit\`a di Roma ``Tor Vergata'',
        Via della Ricerca Scientifi- ca 1,
        00133 Roma, ITALY}
        \thanks{Poster presented by A.Cucchieri.}}
       
\begin{document}

\begin{abstract}
We study the influence of Gribov copies on gluon and ghost
propagators in lattice Landau gauge. For the gluon
propagator, Gribov noise seems to be of the order of magnitude
of the numerical accuracy. On the contrary, for the ghost
propagator, Gribov noise is clearly observable, at least in the
strong-coupling regime.
We also observe, in the limit of large lattice volume,
a gluon propagator
decreasing as the momentum goes to zero.
Finally, we introduce an implementation of the method of Fourier
Acceleration which avoids the use of the fast Fourier transform,
being well suited for parallel and vector machines.
We apply it to the case of Landau gauge fixing,
and study its performance on APE computers.
\end{abstract}

\maketitle

\section{GRIBOV NOISE}

In ref.\ \cite{Gr} Gribov
showed that, for non-abelian gauge theory,
the standard gauge-fixing conditions used for perturbative
calculations do not fix the gauge
fields uniquely. The existence of these {\em Gribov
copies} does not affect the results from perturbation theory, but
their elimination
could play a crucial role for non-perturbative features of
these theories.

In lattice gauge theories
gauge fixing is, in principle,
not required. However, because of asymptotic freedom, the continuum
limit is the weak-coupling limit, and a weak-coupling expansion
requires gauge fixing. Thus, one is
led to consider gauge-dependent quantities on the lattice as well.
Unfortunately gauge fixing on the lattice is
afflicted by the same problem of Gribov copies encountered in the
continuum case \cite{MPR}.

In order to get rid of Gribov copies
the physical configuration space has to be identified with
the so-called {\em fundamental modular region} $\Lambda$, which
is defined (in the continuum) as the set of {\em absolute} minima 
of the functional \cite{STSF}
\be
E_{A}[ g ] \equiv \frac{1}{2}\,\sum_{\mu\mbox{,}\,a}\,
\int\,d^{4}x\,
\left\{\,\left[\,A^{(g)}\,\right]^{a}_{\mu}(x)\,\right\}^{2}
\,\mbox{.}
\label{eq:Econt}
\ee
Similarly, on the lattice, we can eliminate
Gribov copies
looking for the absolute minimum of the
functional ${\cal E}_{U}[ g ]$ ({\em
minimal Landau gauge}) \cite{Z2}
\be
{\cal E}_{U}[ g ] \equiv
\frac{1}{8\,V} \sum_{\mu\mbox{,}\, x} \, \tr \,
             \left[ \, \1 \, - \, U_{\mu}^{(g)}(x) \,
   \right]
\label{eq:Etomin}
\;\mbox{.}
\ee

Given the appearance of Gribov copies in numerical
studies, we need to understand their influence
({\em Gribov noise})
on the evaluation of gauge-dependent quantities.
To this end we compare the results for gluon and
ghost propagators using two different averages \cite{paper1}:
the average considering only the absolute
minima (denoted by ``am''), which should
give us the result in the minimal Landau gauge;
and the average considering
only the first gauge-fixed gauge copy generated
for each configuration (denoted by ``fc''). The latter
average is the result that we would obtain if Gribov noise 
were not considered.

\begin{figure}[htb]
\psfig{figure=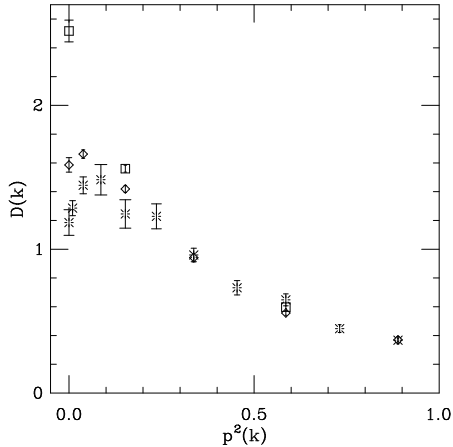,height=2.8in}
\vspace{-1.5cm}
\caption{\protect\small
Plot of the
three-momentum-space gluon
propagator $D(k)$ (``fc''-data), as a function of
the square of the lattice momentum $p^{2}(k)$.
Data correspond to
$V = 16^{3}$ ($\Box$), $V = 32^{3}$ ($\Diamond$)
and $V = 64^{3}$ ($\ast$), at
$\beta = 5.0$.
Error bars are one standard deviation.
Averages are taken over $40$ gauge-fixed
configurations.}
\label{fig:gluon3d}
\vspace{-0.5cm}
\end{figure}
\normalsize

Our data \cite[Table 2]{paper1}
show absence of Gribov noise for the
gluon propagator. In fact, data
corresponding to the minimal Landau gauge (absolute
minima) are in complete agreement, within
statistical errors, with those obtained in a generic
Landau gauge (average ``fc'').
This happens even at $\beta = 0$, where the number of Gribov
copies is very large and
Gribov noise, if present, is more easily
detectable. On the contrary,
a nonzero Gribov noise for the ghost propagator can be
clearly observed \cite[Table 3]{paper1}.
In particular, data corresponding to the
absolute minima (average ``am'') are
constantly smaller than or equal to the corresponding
``fc''-data. This effect is small but clearly detectable
for the values of $\beta$ in the strong-coupling region.
(This was not observed at $\beta = 2.7$.
However, at this value of $\beta$ almost no Gribov copies were
produced, even for a lattice volume $V = 16^4$, and therefore
we cannot expect a difference between the two sets of data.)
This result can been qualitatively explained
\cite{paper1}.
As for the infrared behavior of these two propagators, the data
for the ghost propagator show a pole ``between'' the
zeroth-order perturbative behavior $p^{-2}$ --- valid at
large momenta ---
and the $p^{-4}$ singularity predicted in \cite{Z2},
but in agreement with the pole $p^{-2(1+s)} (0 < s < 1)$
recently obtained in ref.\ \cite{newpaper}.
For the gluon propagator the data show, in the 
{\bf strong-coupling} regime,
a propagator decreasing as the momentum goes to zero.
This anomalous behavior, predicted in \cite{Gr,Z2,Z1}, is still
observable at $\beta = 1.6$, if large volumes are considered
\cite{paper2}.
This result is also observable
in the {\bf scaling region}
in the three-dimensional case, and in the limit of large
lattice volume (see Fig.\ \ref{fig:gluon3d}).
Finally,
the behavior of the
zero three-momentum-space gluon propagator is strongly
affected by the zero-momentum modes of the gluon field
\cite[Fig.\ 2]{paper1},
as predicted in \cite{Mitr}.

\section{FOURIER ACCELERATION}

In order to minimize the functional
${\cal E}_{U}[ g ]$ defined in eq.\ \reff{eq:Etomin},
and to reduce {\em critical slowing-down}, we can
use the Fourier accelerated algorithm \cite{Fourier,gfix123}.
With this algorithm the update is given by
$g^{(new)}(x) \equiv R(x) \, g^{(old)}(x)$, where
\be
R(x) \, \propto \, \1 - {\widehat F}^{-1}\left[
         \, \frac{\alpha}{p^{2}(k)} \,
                {\widehat F}
 \left( \nabla \cdot A^{\left( g \right)} \right)
\right](x)
\;\mbox{.}
\ee
Here ${\widehat F}$ is a Fourier transform,
$\alpha$ is a tuning parameter,
$p^{2}(k)$ is the square of the lattice momentum, and
$\nabla \cdot A$ is the lattice divergence of the
gluon field $A_{\mu}$.
However, this algorithm is of
difficult implementation on parallel machines,
due to the use of the fast Fourier transform (FFT).

Let us notice that
$ {\widehat F}^{-1}\,p^{-2}(k)\,
{\widehat F}
\,=\,\left(\,- \Delta\,\right)^{-1} $,
where $\Delta$ is the lattice Laplacian operator.
Thus, the FFT can be avoided by inverting
$\Delta$ using an algorithm that requires the same computational
work (i.e.\ $V \log N$), such as a multigrid (MG) algorithm
with W cycle and piecewise-constant interpolation.
At the same time, using MG,
we can reduce the computational work with
a good initial guess for the solution, and we can choose the accuracy
of the solution. (With FFT the accuracy is fixed
by the precision used in the numerical code.)
We note that the tuning parameter
$\alpha$ is usually fixed with an accuracy of
a few percent, and thus the inversion of
$\Delta$ should not require a very high accuracy either.

We started our simulations
on an IBM RS-6000/340 workstation.
We tested different types of
multigrid cycles: $\gamma= 0$ (Gauss-Seidel update),
$\gamma= 1$ (V cycle) and
$\gamma= 2$ (W cycle).
We see that MG with $\gamma = 2$, two relaxation sweeps 
on each grid, a minimum of two full multigrid sweeps for each
version of $\Delta$, and an accuracy of $10^{-6}$, is
equivalent to an FFT algorithm \cite{future}.

%
\begin{table}[htb]
\begin{center}
\vspace{-0.5cm}
\begin{tabular*}{7.5cm}{cccc}
\hline
algorithm & $ V $ & GF-sweeps & CPU-time \\
\hline
%
FFT  & $ 4^4 $ & $13.6 \pm 0.3$ & $36 \pm 1$ \\ \hline
MG   & $ 4^4 $ & $13.7 \pm 0.3$ & $65 \pm 3$ \\ \hline
\hline
FFT  & $ 8^4 $ & $16.1 \pm 0.3$ & $972 \pm 16$ \\ \hline
MG   & $ 8^4 $ & $16.1 \pm 0.3$ & $1381 \pm 49$ \\ \hline
\hline
FFT  & $ 16^4 $ & $20.0 \pm 0.4$ & $25254 \pm 560$ \\ \hline
MG   & $ 16^4 $ & $19.9 \pm 0.4$ & $33797 \pm 1114$ \\ \hline
\end{tabular*}
\parbox{7.5cm}{
\vspace{0.2cm}
\caption{\label{tab:accuracy}
\vskip -0.8cm
\hskip 1.25cm
\protect\small
: Comparison between the FFT and MG algorithms at $\beta = \infty$.}
}
\vspace{-1cm}
\end{center}
\end{table}
\normalsize

In Table \ref{tab:accuracy} we report results obtained
at $\beta = \infty$
for the FFT algorithm, and for MG.
Clearly, the two algorithms have a similar performance,
showing a number of gauge-fixing sweeps increasing
logarithmically with the lattice size
$N$, and the CPU-time increasing as $N^4 \log N$.
We also did a test at $\beta = 2.2$ (see Table \ref{tab:finiteb}).
Again, MG is equivalent to the FFT algorithm.

\begin{table}[htb]
\begin{center}
\vspace{-0.8cm}
\begin{tabular*}{7.5cm}{cccc}
\hline
algorithm & $ V $ & GF-sweeps & CPU-time \\
\hline
%
FFT  & $ 8^4 $ & $333 \pm 27$ & $19998 \pm 1236$ \\ \hline
MG   & $ 8^4 $ & $312 \pm 25$ & $18722 \pm 1496$ \\ \hline
\end{tabular*}
\parbox{7.5cm}{
\vspace{0.2cm}
\caption{\label{tab:finiteb}
\vskip -0.8cm
\hskip 1.25cm
\protect\small
: Comparison between the FFT and MG algorithms at $\beta = 2.2$.}
}
\vspace{-1cm}
\end{center}
\end{table}

\normalsize
In order to parallelize,
the idea is to use as the coarsest grid for the
multigrid algorithm a grid with volume equal to or
larger than the number of nodes of the parallel machine.
For example,
for an APE100 computer with $8^3$ nodes we implemented
MG with the coarsest grid $8^4$.
Then, on the coarsest grid, we can use a
Gauss-Seidel relaxation if its volume is
small. Otherwise we can use a Conjugate
Gradient algorithm to relax the solution. (This
combination MG+CG has been used in the past to
accelerate MG on vector machines \cite{vector}.)
In this way, the computational work for the
inversion of $\Delta$ still
increases as V$\log$N,
provided that we keep fixed the size of the
coarsest grid.
We tested this combination first on a workstation
for a $16^4$ lattice with coarsest grid $8^4$ at
$\beta = \infty$, performing two
CG-sweeps when relaxing on the
coarsest grid. We obtained $20.0 \pm 0.4$
for the GF-sweeps, and $46281 \pm 1732$ for the
CPU-time. Therefore (see Table \ref{tab:accuracy})
the performance of
this gauge-fixing algorithm is essentially
equivalent to that of FFT and MG.

Similarly, the performance at 
$\beta = 2.2$,
for an $8^4$ lattice with coarsest grid $4^4$,
is comparable to the performance of the
FFT and MG algorithms (see Table \ref{tab:finiteb}):
we obtained $314 \pm 25$ for the GF-sweeps, and
$23599 \pm 1884$ for the CPU-time.

Finally, we implemented the MG+CG algorithm on an APE100
computer comparing its performance
with a standard overrelaxation (OVE)
and an unaccelerated local algorithm (the so-called
Los Alamos algorithm, LOS) \cite{gfix123}.
The number of gauge-fixing sweeps obtained,
at $\beta = \infty$ and for
lattice volume $16^4$, was $131 \pm 3$ for LOS,
$34.8 \pm 0.5$ for OVE, and $ 16.4 \pm 0.1 $
for MG+CG.
Clearly the MG+CG algorithm is able to reduce the
number of gauge-fixing sweeps compared to the two 
local algorithms.

We plan to extend the tests on APE computers to larger
lattice volumes.

\end{document}